%% file: Lya-CR7.tex
\documentclass[a4paper,fleqn,usenatbib]{mnras}

\usepackage[T1]{fontenc}
\usepackage{ae,aecompl}

\usepackage{graphicx}   
\usepackage{amsmath}    
\usepackage{bm}       
\usepackage{amssymb}    
\usepackage{wasysym}  
\usepackage{txfonts}
\usepackage{etoolbox}
\makeatletter 
  \patchcmd{\NAT@citex}
    {\@citea\NAT@hyper@{%
      \NAT@nmfmt{\NAT@nm}%
      \hyper@natlinkbreak{\NAT@aysep\NAT@spacechar}{\@citeb\@extra@b@citeb}%
      \NAT@date}}
    {\@citea\NAT@nmfmt{\NAT@nm}%
    \NAT@aysep\NAT@spacechar\NAT@hyper@{\NAT@date}}{}{}

  \patchcmd{\NAT@citex}
    {\@citea\NAT@hyper@{%
      \NAT@nmfmt{\NAT@nm}%
      \hyper@natlinkbreak{\NAT@spacechar\NAT@@open\if*#1*\else#1\NAT@spacechar\fi}%
        {\@citeb\@extra@b@citeb}%
      \NAT@date}}
    {\@citea\NAT@nmfmt{\NAT@nm}%
    \NAT@spacechar\NAT@@open\if*#1*\else#1\NAT@spacechar\fi\NAT@hyper@{\NAT@date}}
    {}{}
\makeatother

\newcommand\Msun{\text{M}_{\astrosun}} 
\newcommand\colt{\text{COLT}} 
\newcommand\HI{{H\,\textsc{i}}} 
\newcommand\HII{{H\,\textsc{ii}}} 
\newcommand\HeI{{He\,\textsc{i}}} 
\newcommand\HeII{{He\,\textsc{ii}}} 

\title[Evidence for a DCBH in the Ly$\alpha$ source CR7]{Evidence for a direct collapse black hole in the Lyman $\bm{\alpha}$ source CR7}

\author[A.\ Smith et al.]{
  Aaron~Smith,$^1$\thanks{E-mail: \href{mailto:asmith@astro.as.utexas.edu}{asmith@astro.as.utexas.edu}}
  Volker~Bromm$^1$ and
  Abraham~Loeb$^2$
  \\
  $^1$Department of Astronomy, The University of Texas at Austin, Austin, TX 78712, USA \\
  $^2$Department of Astronomy, Harvard University, 60 Garden Street, Cambridge, MA 02138, USA
}

\date{Accepted 2016 May 10. Received 2016 May 10; in original form 2016 February 24}

\pubyear{2016}

\begin{document}
\label{firstpage}
\pagerange{\pageref{firstpage}--\pageref{lastpage}}
\maketitle

\begin{abstract}
  Throughout the epoch of reionization the most luminous Ly$\alpha$ emitters are capable of ionizing their own local bubbles. The CR7 galaxy at $z \approx 6.6$ stands out for its combination of exceptionally bright Ly$\alpha$ and \HeII\ 1640~\AA\ line emission but absence of metal lines. As a result CR7 may be the first viable candidate host of a young primordial starburst or direct collapse black hole. High-resolution spectroscopy reveals a +160~km\,s$^{-1}$ velocity offset between the Ly$\alpha$ and \HeII\ line peaks while the spatial extent of the Ly$\alpha$ emitting region is $\sim 16$~kpc. The observables are indicative of an outflow signature produced by a strong central source. We present one-dimensional radiation-hydrodynamics simulations incorporating accurate Ly$\alpha$ feedback and ionizing radiation to investigate the nature of the CR7 source. We find that a Population~III star cluster with $10^5$~K blackbody emission ionizes its environment too efficiently to generate a significant velocity offset. However, a massive black hole with a nonthermal Compton-thick spectrum is able to reproduce the Ly$\alpha$ signatures as a result of higher photon trapping and longer potential lifetime. For both sources, Ly$\alpha$ radiation pressure turns out to be dynamically important.
\end{abstract}

\begin{keywords}
  galaxies: formation -- galaxies: high-redshift -- cosmology: theory.
\end{keywords}


\input{text/Introduction}
\input{text/Modeling}
\input{text/Simulation}
\input{text/Conclusions}


\bibliographystyle{mnras}
\bibliography{biblio}

\bsp 
\label{lastpage}
\end{document}

%% file: text/Introduction.tex
\section{Introduction}
\label{sec:introduction}
Before the formation of the first stars and galaxies the cold, expanding Universe was in a state known as the cosmic dark ages. Eventually, gravitational instabilities seeded the growth of dark matter minihaloes with typical mass $M_\text{h} \sim 10^{5-7}~\Msun$, hosting a generation of massive Population~III (Pop~III) stars with a primordial gas composition. The first stars began the process of reionizing the Universe. Over time several of these haloes merge to form atomic cooling haloes with typical mass $M_\text{h} \sim 10^{7-9} \Msun$ by $z \sim 10$, corresponding to the first \textit{bona fide} galaxies \citep{Bromm_Yoshida_2011,Loeb_Furlanetto_2013}. By this time star formation likely consisted of both Pop~III and metal enriched Pop~II stars \citep{Jeon_2015}, although in principle pockets of pristine gas could persist to later redshifts \citep{Pallottini_2014,Ma_2015}. In addition to chemical evolution, the first galaxies experienced significant radiative and mechanical feedback \citep{Pawlik_2013}. Understanding the formation of the first galaxies and their subsequent evolution provides important insights on the remainder of cosmic history and the transition to present-day galaxies.

The high-$z$ observational frontier is constrained by the ability of current infrared technologies to detect increasingly faint and distant galaxies. However, the upgraded capabilities of the \textit{Hubble Space Telescope}~(\textit{HST}) in combination with large aperture ground-based observatories has allowed deep broadband photometry and spectroscopic confirmation of a number of galaxies at $z > 6.5$ \citep{Bouwens_2011,Finkelstein_2013,Oesch_van_Dokkum_2015,Stark_2015,Zitrin_2015}. As predicted by \citet{Partridge_Peebles_1967}, proto-galactic star formation would produce powerful intrinsic Ly$\alpha$ luminosities, although the increasing neutral hydrogen fraction in the diffuse intergalactic medium~(IGM) significantly suppresses Ly$\alpha$ transmission for sources at $z > 6.5$ \citep{Dijkstra_2014,Mesinger_2015}. Still, galaxies with strong Ly$\alpha$ emission are more likely to be detected with substantial outflows or clustered within locally ionized bubbles and web-like morphologies \citep{Jensen_2013,Kakiichi_2015,Smith_2015}. Despite attenuation from the neutral IGM one might hope to infer properties of the emission source, galaxy, and IGM based on the altered Ly$\alpha$ line profile, physical size of the Ly$\alpha$ emitting region, continuum spectral energy distribution~(SED), and relative strength of Ly$\alpha$, H$\alpha$, and \HeII\ 1640~\AA\ nebular emission lines \citep{Schaerer_2002,Johnson_Greif_2009,Pawlik_2011}.

Recent follow-up observations of the Ly$\alpha$ emitter ``CR7'' (COSMOS redshift 7) at $z = 6.6$ by \citet{Sobral_2015} confirmed the presence of strong \HeII\ emission with no detection of metal lines from the UV to the near infrared \citep[see also][]{Matthee_2015}. As the most luminous Ly$\alpha$ emitter at $z > 6$, the CR7 source may be the first detection of a massive ($\gtrsim 10^7~\Msun$) Pop~III star cluster. Deep \textit{HST} imaging reveals three clumps (A, B, and C) in close proximity ($\sim 5$~kpc). Component A dominates the rest-frame UV and is also coincident with both the \HeII\ emission and the centre of the Ly$\alpha$ emitting region. Clumps B and C are much redder and likely dominate the mass of the system. Because a normal stellar population fails to reproduce the excess Ly$\alpha$ and \HeII\ luminosities, \citet{Sobral_2015} suggest a composite spectrum in which clump A is fit by a pure Pop~III SED and clumps B+C are comprised of enriched populations. If CR7 does indeed host a Pop~III star cluster with a top-heavy initial mass function~(IMF) then we are likely viewing a rare starburst with an age of less than a few Myr \citep{Pallottini_2015,Hartwig_2015,Visbal_CR7_2016}.

On the other hand, the Ly$\alpha$ emission from CR7 may also be due to accretion onto a ($\sim 10^{5-6}~\Msun$) black hole. At this redshift the black hole growth time would have been limited to perhaps $\lesssim 500$~Myr providing a strong case for so-called ``direct collapse'' black holes (DCBHs) \citep{Bromm_Loeb_2003}, especially given the existence of supermassive black holes~(SMBHs) of a few billion solar masses observed in $z > 6$ quasars \citep{Fan_2006}. The DCBH formation scenario is explored in detail for the CR7 Ly$\alpha$ emitter with the models of \citet{Pallottini_2015}, \citet{Agarwal_2015}, and \citet{Smidt_CR7_2016}. Furthermore, \citet{Hartwig_2015} use merger tree models to statistically argue against a Pop~III starburst explanation based on their short lifetimes and earlier metal enrichment. Either way, the CR7 enigma is that a massive system was able to maintain a low-metallicity environment down to $z = 6.6$. However, under optimistic assumptions a DCBH is a likely outcome for pristine haloes exposed to strong Lyman-Werner radiation \citep{Hartwig_2015}. Despite the theoretical uncertainties regarding its nature, the CR7 source has tantalizing significance as it opens up additional possibilities for direct observational constraints on first galaxy and quasar formation.

The remainder of this paper is devoted to a discussion of Ly$\alpha$ radiative transfer and radiation pressure in the context of the CR7 source. Stellar feedback has been observed to drive molecular gas outflows in the SDSS\,J0905+57 (redshift $z = 0.712$) compact starburst galaxy \citep{Geach_2014}, and we expect a similar mechanism to operate for Ly$\alpha$ radiation pressure in high-$z$, low metallicity environments \citep{Dijkstra_Loeb_2008,Dijkstra_Loeb_2009}. For a discussion of Ly$\alpha$ feedback in a more general context see \citet{Smith_2016}. In the case of the CR7 Ly$\alpha$ emitter, high-resolution spectroscopy has revealed a +160~km\,s$^{-1}$ velocity offset between the Ly$\alpha$ and \HeII\ line peaks, which is indicative of an outflow signature and/or Ly$\alpha$ scattering in the IGM \citep{Sobral_2015}. A redshifted Ly$\alpha$ line is common in Lyman-$\alpha$ emitting galaxies at $z \lesssim 3$ \citep{Steidel_2010,Erb_2014,Hashimoto_2015,Yang_2016}, nonetheless it is particularly interesting to consider Ly$\alpha$ radiation pressure assisted outflows in the context of CR7. An additional constraint from CR7 (clump A) is the spatial extent of the Ly$\alpha$ emitting region ($\sim 16$~kpc physical). We therefore examine the effect of radiative feedback on these Ly$\alpha$ observables, including Ly$\alpha$ photon trapping and ionizing radiation. In Section~\ref{sec:modeling_CR7} we introduce our simulation model and methodology. In Section~\ref{sec:simulation_results} we present the radiation hydrodynamics and Ly$\alpha$ radiative transfer results. Finally, in Section~\ref{sec:conclusion} we provide a summary and conlusions.

%% file: text/Modeling.tex
\section{Modeling the CR7 system}
\label{sec:modeling_CR7}

\subsection{Basic observational properties}
\label{subsec:basic_observational_properties}
The luminous Ly$\alpha$ emitter CR7 at $z \approx 6.604^{+0.001}_{-0.003}$ stands out for its exceptional Ly$\alpha$ and \HeII\ 1640 \AA\ line emission but absence of metal lines. High-resolution spectroscopy reveals a narrow Ly$\alpha$ line with a FWHM width of 266\,$\pm$15~km\,s$^{-1}$ and luminosity of $L_\alpha > 8.32 \times 10^{43}$~erg\,s$^{-1}$. This is a lower limit because a significant fraction of the Ly$\alpha$ photons could have been `eliminated' from the line of sight due to radiative transfer effects within the galaxy, absorption by dust, anisotropic escape, and scattering by residual \HI\ in the IGM. The \HeII\ 1640 \AA\ line emission on the other hand is free from these complications, therefore we assume the observed luminosity of $L_\text{\HeII} = 1.95 \times 10^{43}$~erg\,s$^{-1}$ accounts for the entire emission. This implies a line ratio of \HeII/Ly$\alpha \lesssim 0.22$ and an ionizing flux ratio of $Q(\text{He}^+) / Q(\text{H}) \lesssim 0.42$, corresponding to a hard blackbody spectrum with effective temperature $T_\text{eff} > 100$~kK \citep{Bromm_Kudritzki_2001,Schaerer_2002}.

The observed CR7 properties provide a strong case for either a young primordial starburst or direct collapse black hole. Given that the halo mass of clump~A is $M_\text{h,A} \approx 3 \times 10^{10}~\Msun$ and the baryonic mass is $M_\text{gas,A} \lesssim 5 \times 10^9~\Msun$ it is not impossible for a massive ($\gtrsim 10^7~\Msun$) Pop~III star cluster to produce the observed \HeII\ emission, although this is pushing standard star formation theory to the extreme. In comparison, Pop~II stars fall short by many orders of magnitude \citep{Hartwig_2015}. Furthermore, the nondetection of metal lines strongly suggests that supernova feedback is not present in CR7. The relatively narrow Ly$\alpha$ line eliminates the possibility of Wolf-Rayet stars and active galactic nuclei~(AGN) as the Ly$\alpha$ source as they produce broader lines with FWHM $\gtrsim 10^3$~km\,s$^{-1}$ \citep{Brinchmann_2008}. The proposal that CR7 is powered by a DCBH does not necessarily imply a similarly broad Ly$\alpha$ line. Quasar lines originate from the broad line region~(BLR) near the vicinity of the black hole, which is made of small gas clouds each contributing a narrow portion of the broad line according to its line-of-sight velocity. The narrow line could be due to recombinations and resonant scattering not associated with a fully developed BLR, or to narrowing during transmission through the IGM.

The +160~km\,s$^{-1}$ velocity offset between the Ly$\alpha$ and \HeII\ line peaks is most likely caused by an outflow within clump~A. It is important to consider the uncertainty on the velocity offset, which is associated with the measured Ly$\alpha$ line width, the redshift estimates, and the spectral resolution. Although the line is broader than the offset it is sharply peaked so the FWHM greatly overestimates the uncertainty. The redshift confidence range $\sigma_z \approx 0.001$ implies a potential redshift uncertainty of at least $\sigma_{\Delta v, z} \approx c \sigma_z / (1+z) \approx 40$~km\,s$^{-1}$. The spectral resolution of DEIMOS/Keck was 0.65~\AA~pix$^{-1}$, which corresponds to an uncertainty of $\sigma_{\Delta v, \text{Keck}} \approx c \Delta \lambda / [\lambda_\alpha (1+z)] \approx 21.1$~km\,s$^{-1}$. The NIR observations with X-SHOOTER/VLT had a resolution of $R \sim 5300$, corresponding to $\sigma_{\Delta v, \text{VLT}} \approx c / R \approx 56.6$~km\,s$^{-1}$ \citep{Sobral_2015}. The spatial extent of the Ly$\alpha$ emitting region is concentrated well within the virial radius of $r_\text{vir,A} \approx 13.3$~kpc. If the offset was caused by scattering in the IGM the Ly$\alpha$ halo would be much more extended \citep{Loeb_Rybicki_1999}. Resonant scattering in a uniform static sphere requires an optical depth at line centre of $\tau_0 \approx 5 \times 10^6 \, T_4^{-1} [\Delta v / (160~\text{km\,s}^{-1})]^3$ to produce the necessary offset, where we have normalized the gas temperature according to $T_4 \equiv T / (10^4~\text{K})$, see the discussion following equation~(34) in \citet{Smith_2015}. Assuming an approximately isothermal density profile, clump~A would have $\tau_0 \gtrsim 10^{11}$ if it remained neutral. Therefore, the effective optical depth must be reduced by a significant velocity gradient and ionization fraction. In this low-metallicity environment the possible outflow mechanisms are hydrodynamical gas pressure, ionizing radiation pressure, and Ly$\alpha$ radiation pressure. We therefore simulate each of these self-consistently in order to determine the effect on the key observables.

\subsection{Numerical methodology}
The simulations in this paper employ the Ly$\alpha$ radiation hydrodynamics (RHD) methods described and tested in \citet{Smith_2016}, which we briefly review now. For computational efficiency and model simplicity we focus on spherically symmetric profiles for this paper. Therefore, we solve the RHD equations in the Lagrangian framework with derivatives denoted by the uppercase differential operator $D(\bullet)/Dt \equiv \upartial(\bullet)/\upartial t + \bmath{v} \bmath{\cdot} \bmath{\nabla}(\bullet)$. The equations for conservation of mass, momentum, and total energy with terms accurate to second order in ($v/c$) are
\begin{align}
  &\frac{D\rho}{Dt} + \rho \bmath{\nabla} \bmath{\cdot} \bmath{v} = 0 \, , \\
  &\frac{D\bmath{v}}{Dt} + \frac{\bmath{\nabla}P}{\rho} = -\bmath{\nabla}\Phi + \bmath{a}_\alpha + \bmath{a}_\gamma \, , \\
  &\frac{D\epsilon}{Dt} + P \frac{D(1/\rho)}{Dt} = \frac{\Gamma - \Lambda}{\rho} \, ,
\end{align}
where $\rho$ is the density, $\bmath{v}$ the velocity, $P$ the pressure, and $\epsilon$ is specific internal energy. Our spherically symmetric hydrodynamics solver is based on the von Neumann-Richtmyer staggered mesh scheme described in \citet{Von_Neumann_1950}, \citet{Mezzacappa_1993}, and \citet{Castor_book_2004}. This has the advantage of adaptive resolution, allowing the grid to follow the motion of the gas. Artificial viscosity is included in order to damp numerical oscillations near shocks. Finally, we assume an ideal gas equation of state so the pressure is specified by $P = (\gamma_\text{ad} - 1) \rho \epsilon$, where in primordial gas we use an adiabatic index of $\gamma_\text{ad} = 5/3$.

The source terms are related to the gravitational potential~$\Phi$, acceleration due to Ly$\alpha$ photons~$\bmath{a}_\alpha$, acceleration from ionizing radiation~$\bmath{a}_\gamma$, and volumetric photoionization heating minus radiative cooling rates, $\Gamma - \Lambda$. In our model the gravitational acceleration includes gas, dark matter, and dark energy components as $\bmath{a}_\text{grav} = -\bmath{\nabla} \Phi = -G M_{<r} \hat{\bmath{r}}/r^2 + H_0^2 \Omega_\Lambda \bmath{r}$, where $M_{<r}$ is the enclosed mass, $H_0$ is the Hubble constant, and $\Omega_\Lambda$ is the dark energy density in units of the present-day critical density. The dark energy component has a negligible effect at these redshifts but is included for completeness. Thus, we adopt physical units for cosmological simulations and directly account for Hubble flow via the initial velocity field. The Ly$\alpha$ radiation pressure is accurately calculated with the Cosmic Ly$\alpha$ Transfer code ($\colt$) according to the Monte-Carlo implementation described in \citet{Smith_2015,Smith_2016}.

Because the ionization state of the gas can greatly affect Ly$\alpha$ radiative transfer we also self-consistently simulate ionizing radiation. We solve the time-dependent transfer equation in an explicitly photon-conserving manner in order to accurately follow the evolution of the ionization front \citep{Abel_1999}. Following \citet{Pawlik_Schaye_2011} and \citet{Jeon_XRB_2014} we collect the photons into three bands according to the ionization energies of \HI, \HeI, and \HeII, or 13.6~eV, 24.6~eV, and 54.4~eV, respectively. Absorption and heating are treated in the grey approximation using the source intensity to compute the photon emission rates~$\dot{N}_{\text{ion},i}$ as well as expectation values for photoionization cross-sections $\langle\sigma_x\rangle_i$ and energy transfer per ionizing photon~$\langle\varepsilon_x\rangle_i$, where the subscript $i \in$~\{1, 2, 3\} denotes the particular band and the subscript $x \in$~\{\HI, \HeI, \HeII\} denotes the species. These quantities completely describe the central source of ionizing radiation in our models. Finally, we also solve the rate equations for a primordial chemistry network consisting of H, H${}^+$, He, He${}^+$, He${}^{++}$, and e$^{-}$ \citep{Bromm_2002}. Reactions affecting these abundances include photoionization, collisional ionization, and recombination while cooling mechanisms also include collisional excitation, bremsstrahlung, and inverse Compton cooling \citep{Cen_1992}. As photoionization can rapidly affect the chemical and thermal state of the gas we employ timestep sub-cycling to accurately follow the evolution.

For the initial simulation setup we adopt an idealized model galaxy with the observed properties of the CR7 system. For simplicity we assume an NFW dark matter halo profile and thereby calculate the dark matter gravitational contribution analytically. Specifically, the dark matter density is $\rho_\text{DM}(r) = \rho_{\text{DM},0} R_\text{S}^3 / [r (R_\text{S} + r)^2]$ with a concentration parameter of $c_\text{NFW} \equiv R_\text{vir} / R_\text{S} \approx 5$, where $R_\text{vir}$ is the virial radius and $R_\text{S}$ is the scale radius. The gas follows an isothermal density profile with $\rho(r) = \rho_\text{vir} (r / R_\text{vir})^{-2}$ until it reaches the background IGM density $\rho_\text{IGM}(z)$. The galactic gas starts off cold and neutral, with an initial temperature set by the cosmic microwave background at $T = 2.725~\text{K}~(1 + z)$. The exact value makes little difference as the gas quickly becomes hot and ionized within the growing \HII\ region. The simulation begins at redshift $z = 6.6$ and continues for $10-100$~Myr.

\subsection{Possible sources}
The delayed Pop~III starburst and massive black hole~(MBH) scenarios are characterized by central emission with different source parameters. Each model is normalized to match the \HeII\ emission rate which we calculate to be $\dot{N}_\text{\HeII} \approx 3.44 \times 10^{54}~\text{s}^{-1}$ \citep{Bromm_Kudritzki_2001,Schaerer_2002}. The specific source parameters for each model considered in this paper are presented in Table~\ref{tab:source_parameters}. The fiducial Pop~III case is calculated from a blackbody spectrum with effective temperature $T_\text{eff} = 100$~kK. We also consider modified scenarios in which (\textit{i}) the Ly$\alpha$ luminosity is boosted by an order of magnitude and (\textit{ii}) the ionizing emission rates are reduced by an order of magnitude as a crude approximation for continuum leakage within a one-dimensional setting. The $\sim 10^5~\Msun$ black hole case is based on the popular DCBH formation model, although our use of the MBH acronym allows for other formation scenarios. We model the MBH according to the nonthermal Compton-thick spectrum presented by \citet{Pacucci_MBH_Spectra_2015}. Specifically, we use the supplemental data provided by the authors to calculate typical ionization rates for their low-density profile, standard accretion scenario at $\lesssim 75$~Myr. The Pop~III and MBH SEDs differ significantly, which is apparent by comparing the photon emission rates, photoionization cross-sections, and energy transfer per ionizing photon in Table~\ref{tab:source_parameters}. The Compton-thick MBH environment reprocesses the broadband spectrum such that only X-ray and non-ionizing photons remain. Note, for simplicity we model the cascade of multiple ionization and heating events experienced by the X-ray photons with a `one-shot' approximation, where all the energy is transferred in a single scattering \citep[e.g.][]{Shull_1985}.

The ionizing photons are responsible for heating the gas within the galaxy and thereby produce a high-velocity shock. The Str{\"o}mgren radius for such a strong source surrounded by the neutral IGM with a background density of $n_\text{\HI}|_{z=6.6} \approx 4.3 \times 10^{-4}~\text{cm}^{-3}$, Case~B recombination coefficient $a_\text{B} \approx 2.59 \times 10^{-13} \text{cm}^3\text{s}^{-1}$, and rate of ionizing photons $\dot{N}_\text{ion} \approx 5 \times 10^{54}~\text{s}^{-1}$ is of order $r_\text{S} \approx [3 \dot{N}_\text{ion}/(4 \pi n_\text{\HI}^2 \alpha_\text{B})]^{1/3} \sim 1$~Mpc. Therefore, the CR7 source is capable of ionizing its own local super bubble. Furthermore, the ionization parameters and resultant state of the gas has a strong impact on the emergent Ly$\alpha$ spectra. The modifications to our fiducial models are designed to explore the impact of source parameters and gain additional physical insight.

\begin{table}
  \caption{Summary of the source parameters for the Pop~III and black hole models. The `boost' and `leak' modifications respectively increase the Ly$\alpha$ luminosity or reduce the ionizing photon rate. The Ly$\alpha$ luminosity is taken to be $L_\alpha = 0.68 h \nu_\alpha (1 - f^\text{ion}_\text{esc}) \dot{N}_\text{ion}$ where $h \nu_\alpha = 10.2$~eV, $f^\text{ion}_\text{esc}$ is the escape fraction of ionizing photons, and $\dot{N}_\text{ion} = \sum \dot{N}_{\text{ion},i}$.}
  \label{tab:source_parameters}
  \begin{tabular}{@{} c cccc @{}}
    \hline
    Model & Pop III & Black Hole & Boost & Leak \\
    \hline
    \vspace{.05cm}
    \;$L_\alpha$ \;[erg\,$s^{-1}$]\; & $6.53 \times 10^{44}$ & $3.82 \times 10^{43}$ & $\times \, 10$ & -- \\
    \;$\dot{N}_{\text{ion},1}$ \;[s$^{-1}$]\; & $2.63 \times 10^{55}$ & $0$                   & -- & $\div \, 10$ \\
    \;$\dot{N}_{\text{ion},2}$ \;[s$^{-1}$]\; & $2.90 \times 10^{55}$ & $0$                   & -- & $\div \, 10$ \\
    \;$\dot{N}_{\text{ion},3}$ \;[s$^{-1}$]\; & $3.44 \times 10^{54}$ & $3.44 \times 10^{54}$ & -- & $\div \, 10$ \\
    \;$\dot{N}_\text{ion}$  \;[s$^{-1}$]\; & $5.88 \times 10^{55}$  & $3.44 \times 10^{54}$  & -- & $\div \, 10$ \\
    \;$\langle\sigma_\text{\HI}\rangle_1$   \;[cm$^2$]\; & $3.00 \times 10^{-18}$ & $0$                    & -- & -- \\
    \;$\langle\sigma_\text{\HI}\rangle_2$   \;[cm$^2$]\; & $5.69 \times 10^{-19}$ & $0$                    & -- & -- \\
    \;$\langle\sigma_\text{\HI}\rangle_3$   \;[cm$^2$]\; & $7.88 \times 10^{-20}$ & $2.61 \times 10^{-23}$ & -- & -- \\
    \;$\langle\sigma_\text{\HeI}\rangle_2$  \;[cm$^2$]\; & $4.67 \times 10^{-18}$ & $0$                    & -- & -- \\
    \;$\langle\sigma_\text{\HeI}\rangle_3$  \;[cm$^2$]\; & $1.48 \times 10^{-18}$ & $7.42 \times 10^{-21}$ & -- & -- \\
    \;$\langle\sigma_\text{\HeII}\rangle_3$ \;[cm$^2$]\; & $1.05 \times 10^{-18}$ & $5.66 \times 10^{-22}$ & -- & -- \\
    \;$\langle\varepsilon_\text{\HI}\rangle_1$   \;[eV]\; & 3.85 & 0   & -- & -- \\
    \;$\langle\varepsilon_\text{\HI}\rangle_2$   \;[eV]\; & 17.5 & 0   & -- & -- \\
    \;$\langle\varepsilon_\text{\HI}\rangle_3$   \;[eV]\; & 48.4 & 579 & -- & -- \\
    \;$\langle\varepsilon_\text{\HeI}\rangle_2$  \;[eV]\; & 8.01 & 0   & -- & -- \\
    \;$\langle\varepsilon_\text{\HeI}\rangle_3$  \;[eV]\; & 38.5 & 728 & -- & -- \\
    \;$\langle\varepsilon_\text{\HeII}\rangle_3$ \;[eV]\; & 7.89 & 550 & -- & -- \\
    \hline
  \end{tabular}
\end{table}

%% file: text/Simulation.tex
\section{Simulation results}
\label{sec:simulation_results}

  \begin{figure}
    \centering
    \includegraphics[width=\columnwidth]{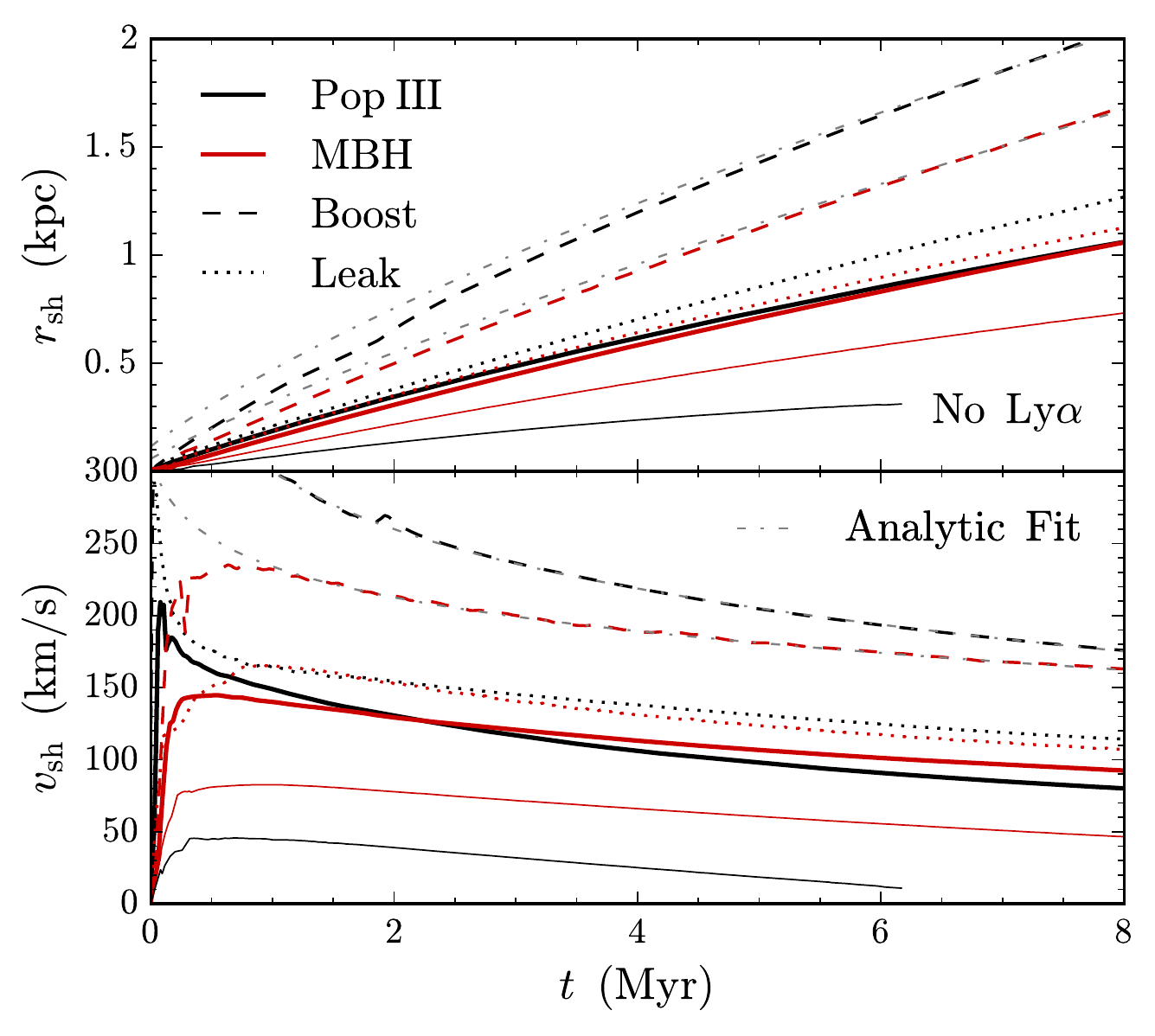}
    \caption{\protect\input{figures/rv/caption}}
    \label{fig:rv}
  \end{figure}

\subsection{Outflow Structure}
Soon after the central source turns on the gas within the galaxy becomes hot and ionized. Eventually a dense shell forms in hydrodynamical response and expands in accordance with conservation of momentum. If we assume the shell continually accumulates mass as it sweeps outward into an isothermal density profile then the mass grows with radius as $m(r) = 4 \pi \int_0^r \rho(r') r'^2 dr' \propto r$. We may extend the analysis by considering power-law profiles with $m(r) = m_0 (r/r_0)^\beta$ where $\beta \in (0,1)$ and the subscript~$0$ denotes a specified moment in time. Therefore, momentum conservation dictates that
\begin{equation} \label{eq:shell_eom}
  \frac{\text{d}(mv)}{\text{d}t} = \frac{\text{d}}{{\text{d}t}}\left[m_0 \left(\frac{r}{r_0}\right)^\beta \dot{r}\right] = F(t) \, ,
\end{equation}
where $F(t)$ is the force on the shell. Examples of the force contributions are from Ly$\alpha$ photons $\sim L_\alpha / c$, gravity $\sim G m(r) M(<r) / r^2 \propto r^{\beta-1}$, and gas pressure $\sim 4 \pi r^2 \rho c_\text{s}^2$ where $c_\text{s}$ is the sound speed. In the case that these forces vary slowly in time or are small compared to the initial kick, Equation~(\ref{eq:shell_eom}) admits the following power-law solution for the shell's motion:
\begin{align} \label{eq:shell_sol}
  & f \equiv 1 + (1+\beta) \frac{(t-t_0)}{r_0} \left( v_0 +\frac{a_0}{2} (t-t_0) \right) \, , \notag \\
  & r(t) = r_0 f^\frac{1}{1+\beta} \qquad \text{and} \qquad v(t) = \left[ v_0 + a_0 (t-t_0) \right] f^\frac{-\beta}{1+\beta} \, ,
\end{align}
where at time~$t_0$ the radius, velocity, and acceleration are respectively $r_0$, $v_0$, and $a_0$. Equation~(\ref{eq:shell_sol}) provides a reasonable description of the evolution at late times, despite the apparent breakdown in the validity of our assumptions early on. The radial expansion of the shells in each of the simulation models is shown in Fig.~\ref{fig:rv}. The shell maintains a $v_\text{sh} \gtrsim 100~\text{km\,s}^{-1}$ outflow throughout the simulation. We also show that Ly$\alpha$ radiation pressure significantly impacts the shell velocity, which is evident in comparison to simulations without Ly$\alpha$ coupling included. Furthermore, as expected, the Ly$\alpha$ feedback is noticeably more effective in the models with a boost in Ly$\alpha$ luminosity or an ionizing continuum leakage.

  \begin{figure}
    \centering
    \includegraphics[width=\columnwidth]{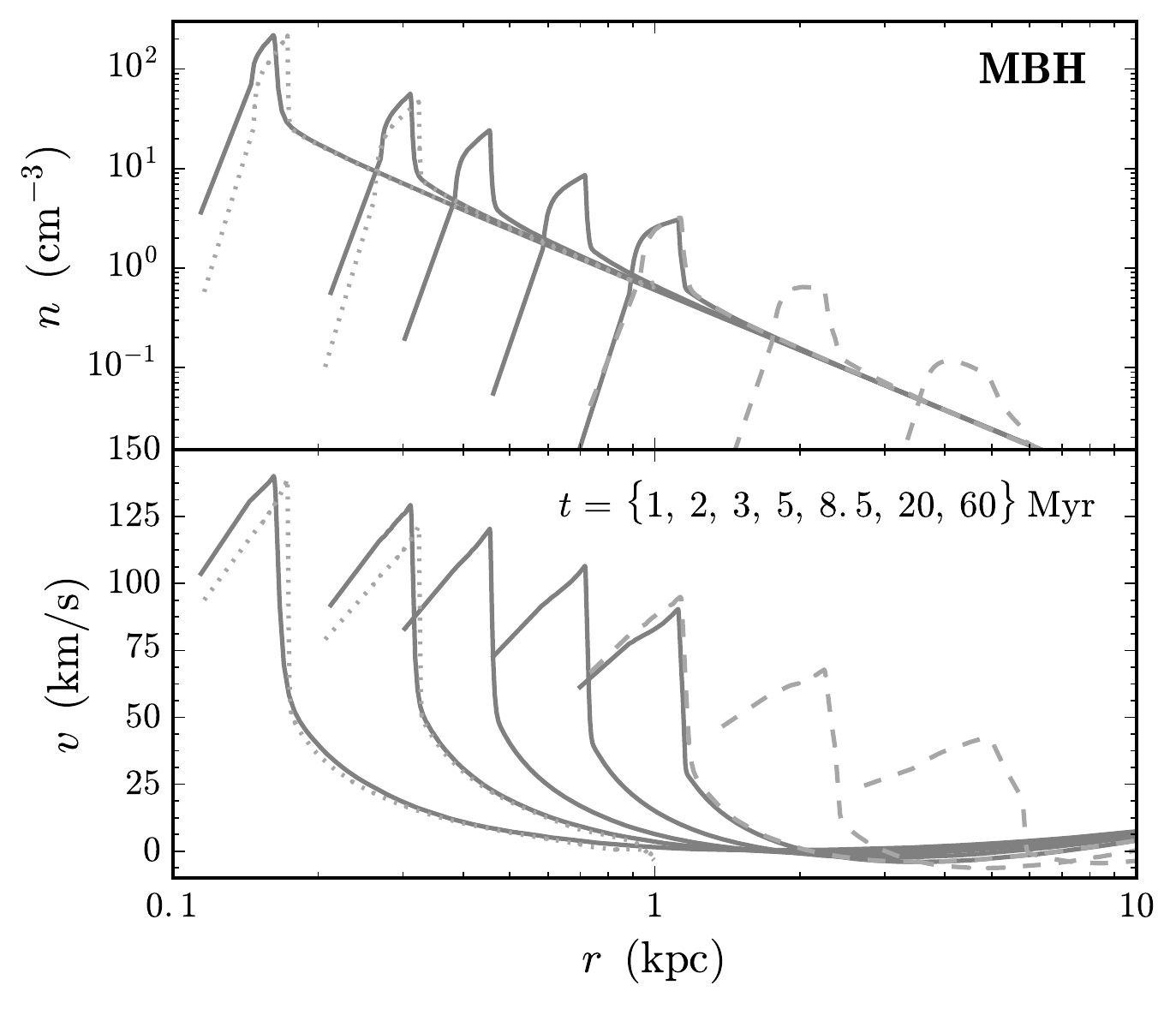}
    \caption{\protect\input{figures/hydro/caption}}
    \label{fig:hydro}
  \end{figure}

\begin{figure}
    \centering
    \includegraphics[width=\columnwidth]{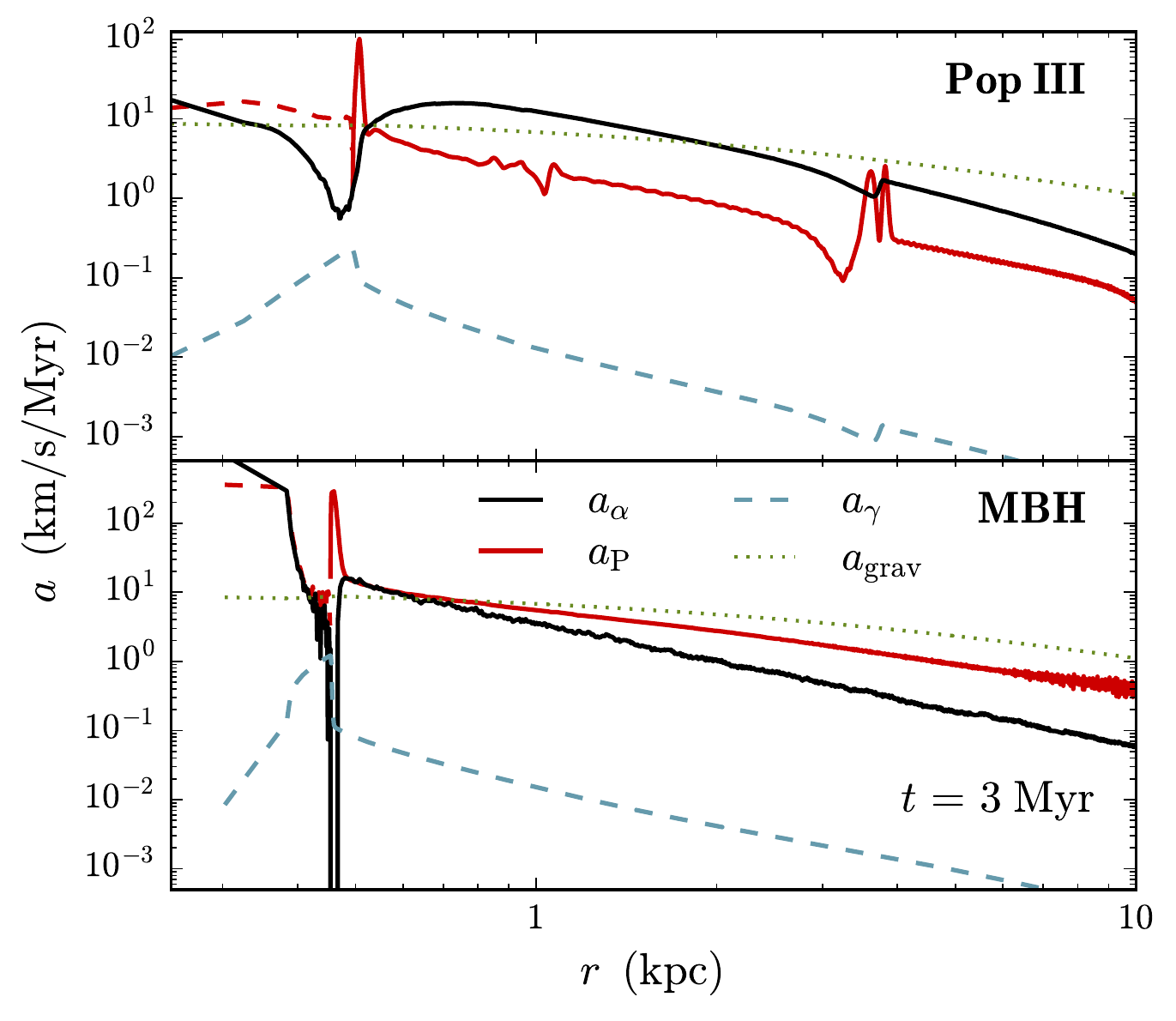}
    \caption{\protect\input{figures/acceleration/caption}}
    \label{fig:acceleration}
  \end{figure}

Similarly, we present radial profiles of the gas number density and velocity at various stages throughout the MBH simulation in Fig.~\ref{fig:hydro}. The evolution of the shell outflow is qualitatively similar in each of the models with slight differences in shape. At late times, e.g. as reflected in the $60$~Myr curve, the sharp shock features naturally dissipate as it continues to propagate into lower density gas. In Fig.~\ref{fig:acceleration} we consider the radial acceleration due to Ly$\alpha$ photons, gas pressure, ionizing radiation, and gravity (dark matter + baryons). The relative importance of Ly$\alpha$ feedback and hydrodynamical pressure is inverted between the Pop~III and MBH cases. In the massive black hole case Ly$\alpha$ radiation undergoes significant trapping within the shell and Ly$\alpha$ feedback is dominant within the shell. However, due to resonant scattering in expanding media the photons emerge redward of line centre and therefore the Ly$\alpha$ force is reduced. One of the predominant features in both models is that the gas pressure is strongly peaked at the expanding shell front. Finally, ionizing momentum transfer roughly traces the neutral gas density and is therefore dynamically unimportant once the gas is ionized.

\subsection{Velocity offset}
\begin{figure}
    \centering
    \includegraphics[width=\columnwidth]{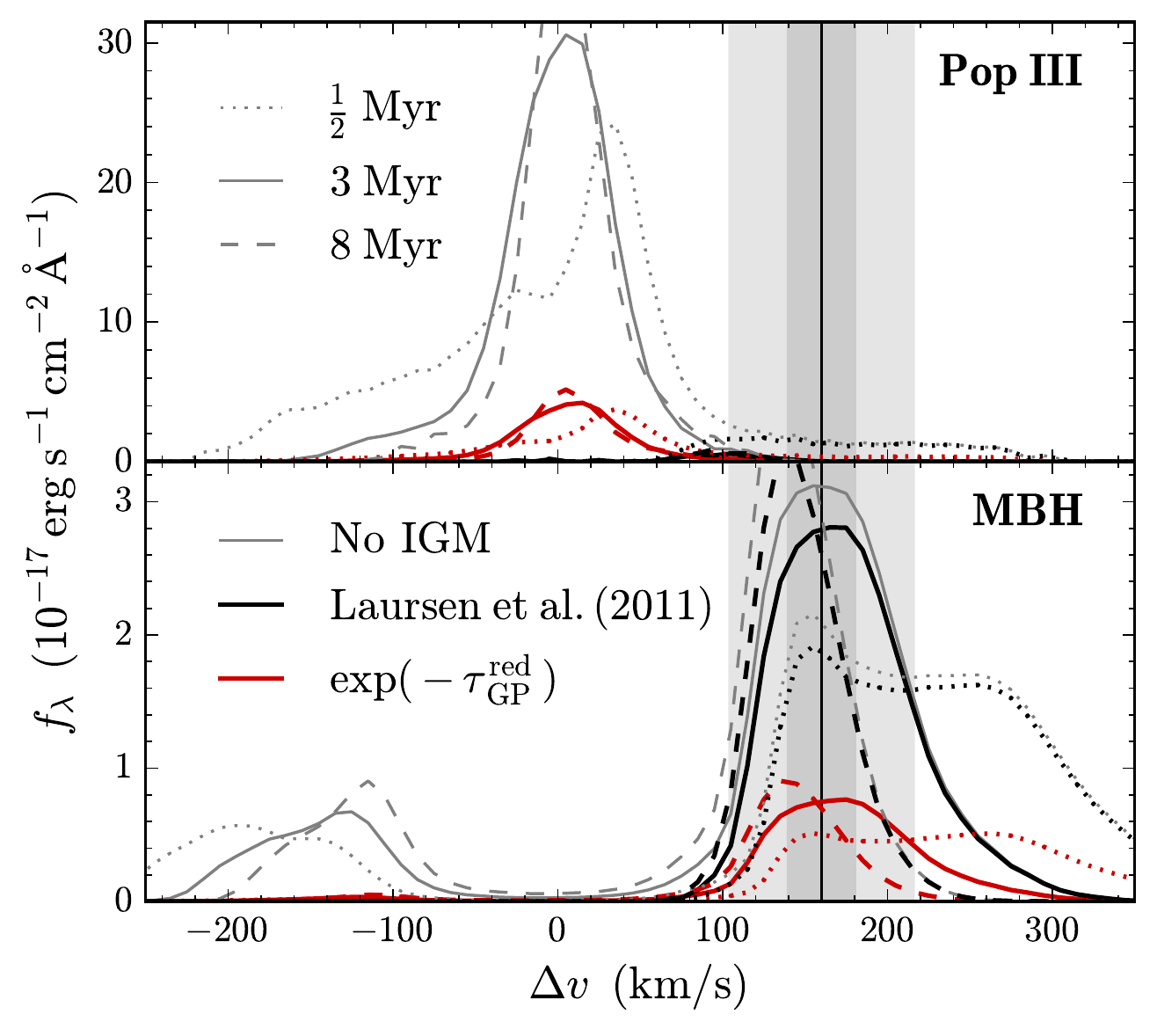}
    \caption{\protect\input{figures/flux/caption}}
    \label{fig:flux}
  \end{figure}

\begin{figure}
    \centering
    \includegraphics[width=\columnwidth]{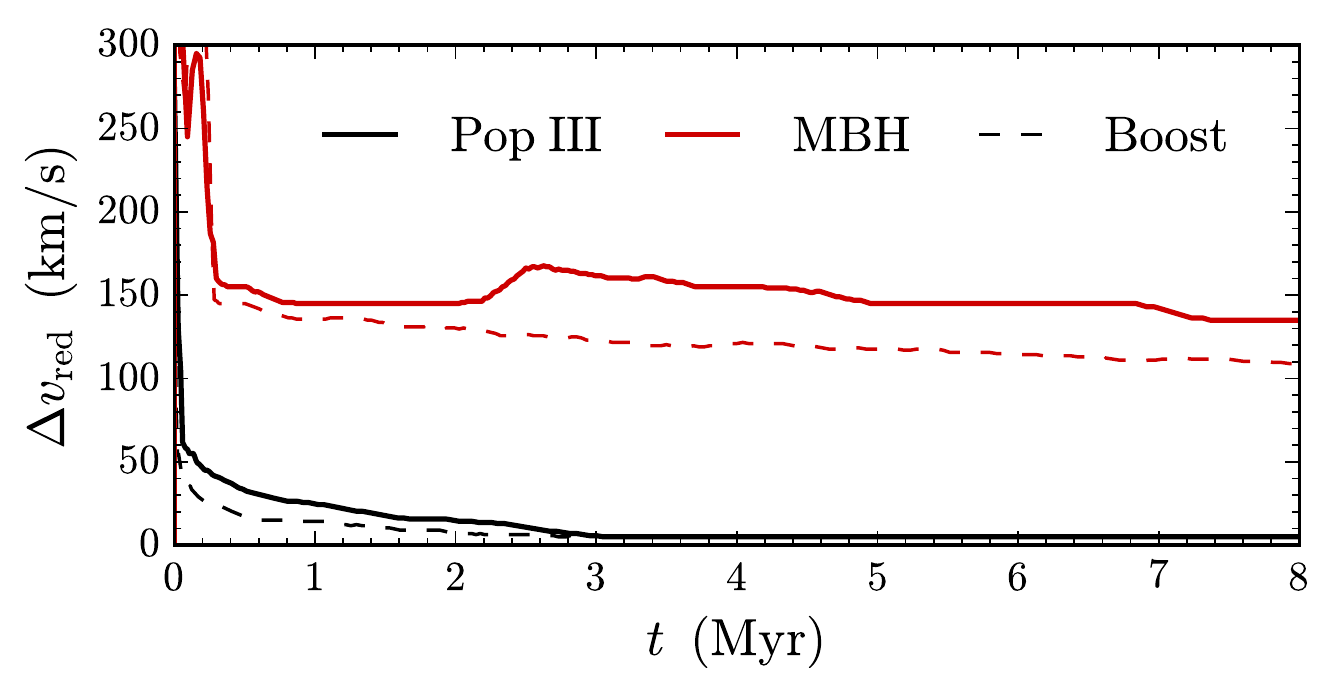}
    \caption{\protect\input{figures/v_offset/caption}}
    \label{fig:v_offset}
  \end{figure}

We now consider whether the models self-consistently reproduce the velocity offset observed between the Ly$\alpha$ and \HeII\ lines. Figure~\ref{fig:flux} shows the typical Ly$\alpha$ line flux for the Pop~III and MBH models from early to late times of $0.5$~Myr, $3$~Myr, and $8$~Myr. Although each simulation environment is highly ionized by this time the two sources yield vastly different spectral profiles. This may be understood in terms of the residual \HI\ within the expanding shell. As seen from Table~\ref{tab:source_parameters}, the Pop~III source has a much higher hydrogen effective ionization cross-section than the black hole emission. In fact, the degree of ionization is such that Ly$\alpha$ photons emerge without a significant systemic velocity offset. On the other hand, the ionizing radiation in the MBH model leaves sufficient \HI\ so that Ly$\alpha$ resonant scattering becomes coupled to the outflow. Therefore, the observed velocity offset provides a powerful hint into the nature of CR7 and provides an argument against a $10^5$~K blackbody radiation source as might be produced from Pop~III stars. In Figure~\ref{fig:v_offset} we show the time evolution of the velocity offset for the red peak of the intrinsic line-of-sight flux, which is fairly constant throughout the simulations demonstrating the robustness of these models.

We use the frequency dependent transmission curve of \citet{Laursen_2011} to account for subsequent scattering of Ly$\alpha$ photons out of the line of sight. Specifically, we use their ``benchmark'' Model~1 curve at $z \approx 6.5$ as described in relation to their figures~2~and~3. The curve is a statistical average of a large number of sightlines ($\gtrsim 10^3$) cast from several hundreds of galaxies through a simulated cosmological volume. Any IGM model should be considered in the context of statistical variations depending on the particular galaxy and sightline. However, for concreteness we multiply their average IGM transmission curve with our intrinsic Ly$\alpha$ spectra when discussing the CR7 observables for the \citet{Laursen_2011} model.

Our second treatment of transmission through the IGM represents a plausible reprocessing of the spectrum based on the analytic prescription of \citet{Madau_Rees_2000}. Ly$\alpha$ photons are removed from the line of sight flux according to their likelihood of undergoing even a single scattering event. In an Einstein-de Sitter universe with a uniform, neutral IGM the optical depth of the red damping wing of the Gunn-Peterson~(GP) trough is
\begin{equation} \label{eq:tau_GP}
  \tau_\text{GP}^\text{red} = \frac{\tau_0(z)}{\upi R_\alpha^{-1}} \left( 1 - \frac{\Delta v}{c} \right)^{3/2} \int_{x_\text{re}}^{x_\text{\HII}} \frac{\text{d}x\;x^{9/2}}{(1-x)^2 + R_\alpha^2 x^6} \, ,
\end{equation}
where $R_\alpha \equiv \Lambda_\alpha \lambda_\alpha / (4 \pi c) \approx 2 \times 10^{-8}$, the optical depth at line centre is $\tau_0(z) \approx 5 \times 10^5\,[(1+z)/7.7]^{3/2}$, and the dimensionless parameters are given by $x_i = (1 - \Delta v/c) (1+z_i) / (1+z)$. The limits of integration are set by a cutoff at the reionization redshift $z_\text{re} \approx 6$ and the size of the ionized bubble around the source via $z_\text{\HII}$ corresponding to a physical radius of $r_\text{\HII} \approx 500$~kpc. Calculations of Ly$\alpha$ observables are based on the intrinsic line of sight flux attenuated by this approximate transmission through the IGM, i.e. $F_{\lambda,\text{obs}} \equiv F_\lambda e^{-\tau_\text{IGM}}$. As illustrated in Fig.~\ref{fig:flux} the observed luminosity may be reduced substantially with a ratio of $L/L_\text{obs} \approx 46$ and $\approx 1.6$ for the Pop~III and MBH cases, respectively, under the \citet{Laursen_2011} IGM model. For the $\tau_\text{GP}^\text{red}$ IGM model the ratio is $L/L_\text{obs} \approx 7.7$ and $\approx 4.7$ for the Pop~III and MBH cases.

\subsection{Spatial extent}
\begin{figure}
    \centering
    \includegraphics[width=\columnwidth]{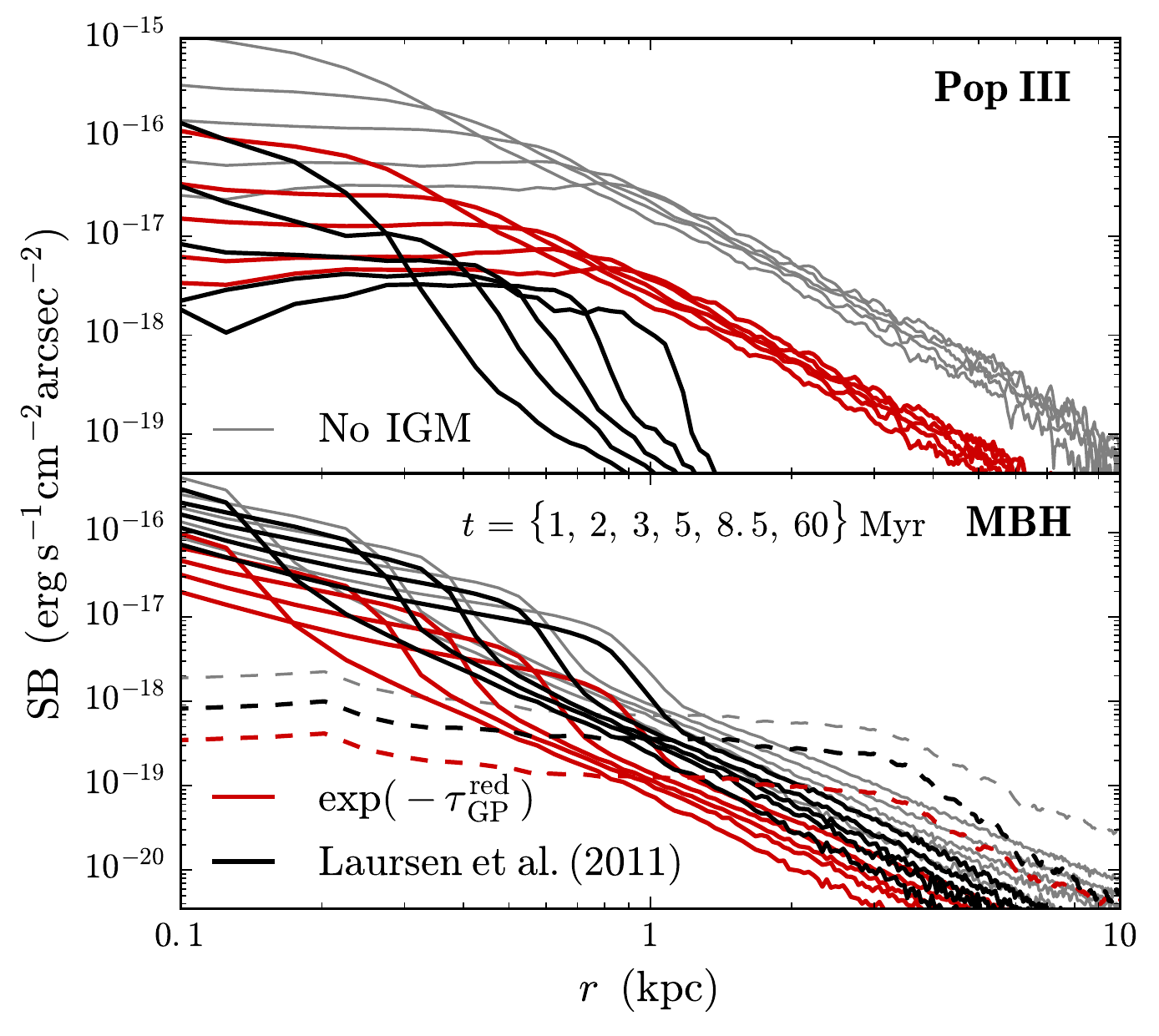}
    \caption{\protect\input{figures/SB_r/caption}}
    \label{fig:SB_r}
  \end{figure}

An additional observational constraint from CR7 is the $\sim 16$~kpc spatial extent of the Ly$\alpha$ emitting region. To examine this we show the evolution of the radial Ly$\alpha$ surface brightness profile in Fig.~\ref{fig:SB_r} for the Pop~III and MBH models. As the shell expands the central intensity is reduced while the outer intensity increases. The drop in brightness at the shell front is much sharper in the black hole model than the Pop~III case due to the more efficient Ly$\alpha$ trapping. The observed surface brightness reflects the estimated transmission through the IGM based on the models of \citet{Laursen_2011} and that of Equation~(\ref{eq:tau_GP}). Photons that scatter toward an observer at larger radii are typically closer to line centre and therefore suffer greater absorption. Specifically, for the MBH model we find an intrinsic to observed ratio of $SB_r/SB_{r,\text{obs}} \approx \{1.2, 2, 7, 20\}$ at $r \approx \{0.05, 1, 10, 20\}$~kpc, respectively, under the \citet{Laursen_2011} IGM model. For the $\tau_\text{GP}^\text{red}$ IGM model the ratio is $SB_r/SB_{r,\text{obs}} \approx \{4, 6.5, 7.5\}$ at $r \approx \{0.05, 1, 20\}$~kpc, respectively.

\begin{figure}
    \centering
    \includegraphics[width=\columnwidth]{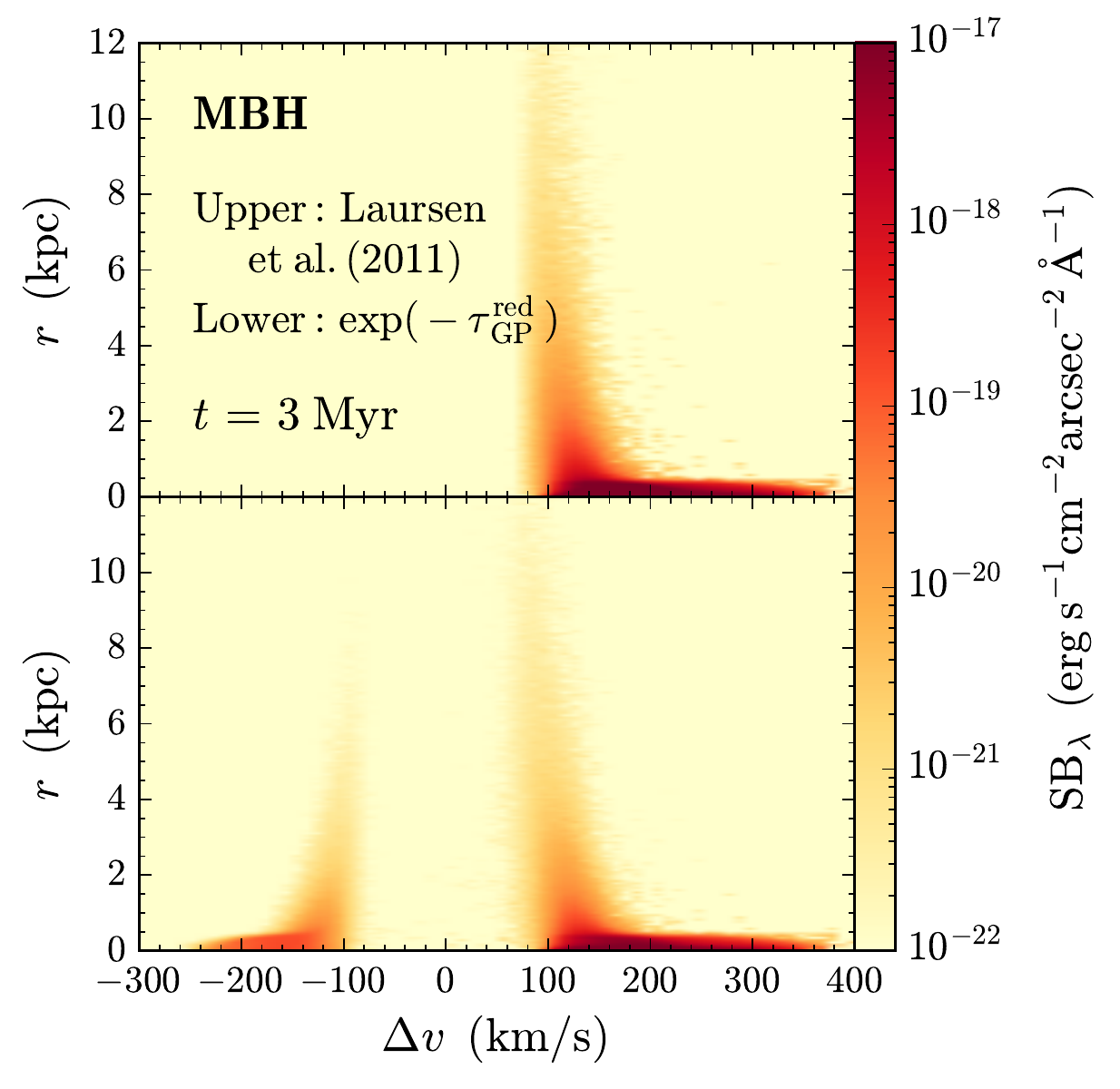}
    \includegraphics[width=\columnwidth]{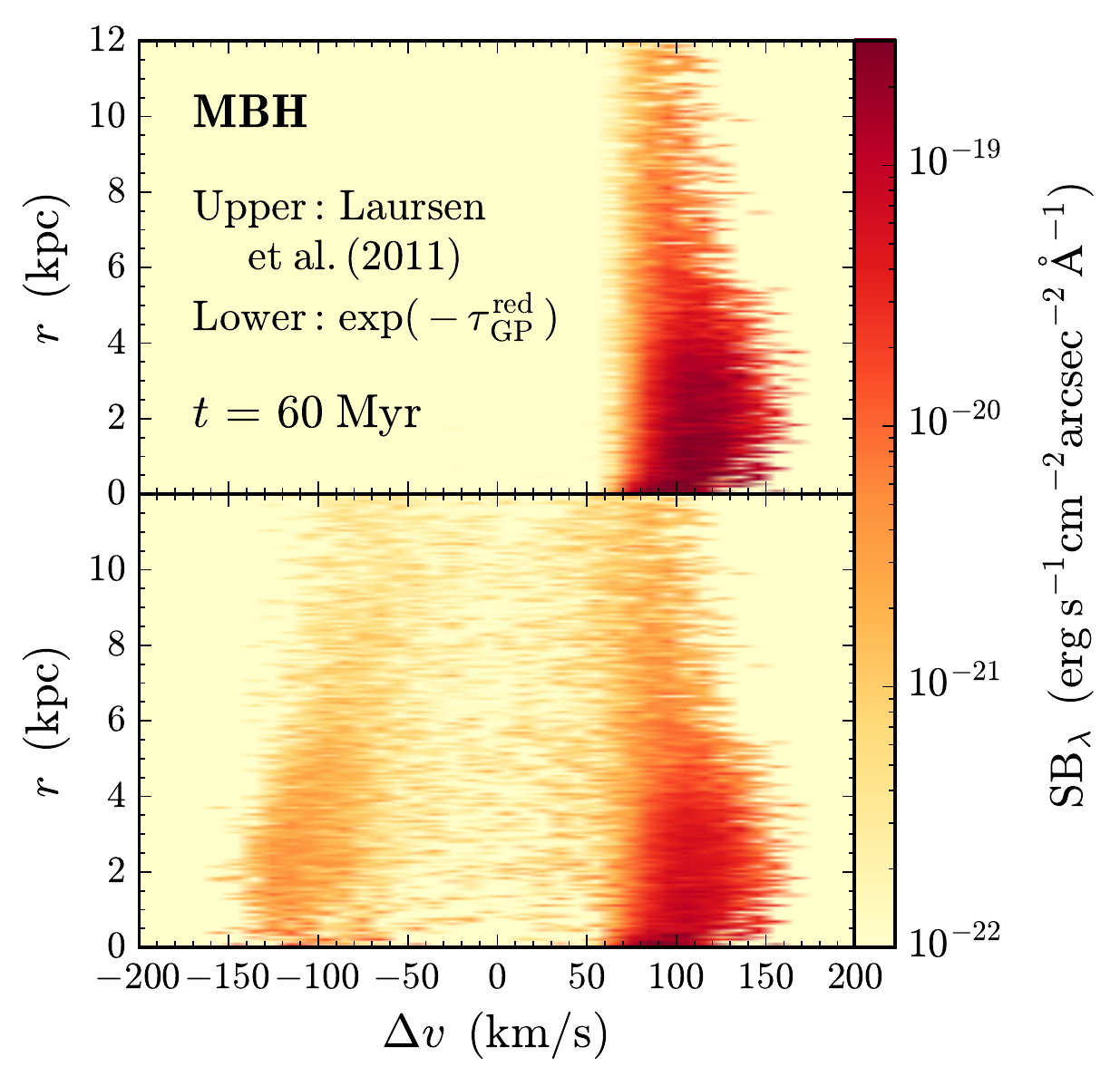}
    \caption{\protect\input{figures/SB_rv/caption}}
    \label{fig:SB_rv}
  \end{figure}

Complementary to the radial surface brightness profile is the spectral line intensity as would be observed along a slit in the radial direction. Figure~\ref{fig:SB_rv} illustrates such a measurement for the MBH model at simulation times of $3$~Myr and $60$~Myr. The images have been corrected for IGM transmission and show the relative importance of the red peak compared to the blue peak which is visible due to the logarithmic scaling. This perspective also demonstrates a slight decrease in the velocity offset at larger radii which is explained by the resonant scattering bias to return to line-centre. The primary difference between the early and late times is the radius of the shell front which by $60$~Myr has advanced to $\sim 4$~kpc and $\sim 8$~kpc in the MBH and MBH Boost models, respectively. At later times the velocity offset decreases, possibly as a result of the lower shell velocity and a more diffuse Ly$\alpha$ trapping environment. Still, all of the aforementioned Ly$\alpha$ features persist in simulations carried out to $120$~Myr. If the shell radius is indicative of the Ly$\alpha$ emitting region then even at $v_\text{sh} \sim 100~\text{km\,s}^{-1}$ it would take at least $50$~Myr to achieve a size of $\sim 10$~kpc in diameter. Although our models allow $v_\text{sh}$ to be a few times larger, we nonetheless find it less probable that the CR7 source is significantly younger than this. Again, such long source durations can more easily be explained within a MBH scenario than a Pop~III starburst.

\subsection{Caveats}
We now discuss the limitations of our simulations and their potential impact on the above results. Uncertainties in the galaxy model parameters and initial conditions may produce slightly different observables. For example, normalizing the black hole flux to the observed \HeII\ emission leads to a much lower Ly$\alpha$ luminosity. For this reason we have explored the ``Boost'' and ``Leak'' models to determine the impact of extra Ly$\alpha$ emission or less ionizing radiation. We also modeled the photon emission as a central point source. While the Pop~III starburst is initially centrally dominated, it is highly unlikely to remain so as star formation proceeds. However, similar to \citet{Dijkstra_Loeb_2008} we expect our final results are not particularly sensitive to the choice for the radial Ly$\alpha$ photon emission.

Another issue is that of numerical resolution. At a halo mass of $M_\text{vir} \approx 3 \times 10^{10}~\Msun$ we achieve Lagrangian mass elements of $\approx 6.6 \times 10^5~\Msun$ which at various times corresponds to $\approx 0.1-2$~pc resolution at the shell front. Improving the resolution by a factor of $\approx 5$ produces very similar results except at the earliest times or innermost radii. Higher resolution leads to slightly faster initial velocities because higher densities allow for more efficient radiation coupling. However, the computational cost may change dramatically based on the number of timesteps with MCRT calculations and the convergence criteria. To be safe we require that between batches of $\approx 1600$ photons the Ly$\alpha$ force has a $\lesssim 1$ per cent relative change in $\gtrsim 99$ per cent of the cells.

Furthermore, distinguishing effects due to radiative transfer within the galaxy and in the IGM can be difficult to disentangle. We have used two IGM transmission models based on cosmological simulations and a simple analytic treatment. Still, the true Ly$\alpha$ transmission could deviate from each of these. See \citet{Dijkstra_2014} for a review containing a detailed discussion and related references. An important consequence of some IGM models is that the observed velocity offset could be mimicked by a broad spectrum or a static double-peaked spectrum in which the flux blueward of the observed +160~km\,s$^{-1}$ offset is suppressed by inflow into the potential well \citep{Loeb_Eisenstein_1995,Barkana_Loeb_2003}. Future observations could potentially distinguish between these scenarios.

We also note that the input spectrum impacts other Ly$\alpha$ observables. For example, in Fig.~\ref{fig:flux} the emergent spectra have FWHM at $0.5$~Myr, $3$~Myr, and $8$~Myr of respectively $\approx 60$~km\,s$^{-1}$, $70$~km\,s$^{-1}$, and $50$~km\,s$^{-1}$ for the Pop~III model and $\approx 180$~km\,s$^{-1}$, $100$~km\,s$^{-1}$, and $60$~km\,s$^{-1}$ for the MBH model. All of these values are significantly narrower than the observed 266\,$\pm$15~km\,s$^{-1}$ FWHM of the Ly$\alpha$ line. This indicates that the width is dominated by unresolved turbulence and three-dimensional radiative transfer effects. As a rough estimate we assume $\text{FWHM} \propto v_\text{eff} \approx ( c_\text{s}^2 + v_\text{turb}^2 )^{1/2} \approx c_\text{s} ( 1 + \text{Ma}^2 )^{1/2} \approx c_\text{s} \text{Ma}$, where Ma is the Mach number. A comparison with the calculated FWHM implies $\text{Ma} \approx 2-5$, which is reasonable but not incorporated into our simulations. We may also compare our results with the observed line ratio of \HeII/Ly$\alpha \approx 0.22$. At 3~Myr with the \citet{Laursen_2011} ($\tau_\text{GP}^\text{red}$) IGM transmission we find \HeII/Ly$\alpha \approx 3.36~(0.37)$ for the Pop~III model and \HeII/Ly$\alpha \approx 0.41~(1.4)$ for the MBH model. All of these are high by a factor of $\gtrsim 2$, which is difficult to account for in the Pop~III case but the MBH case may allow for more Ly$\alpha$ photons than inferred by the Compton-thick spectrum. The `boost' modification explores the effect of an order of magnitude overcorrection to this ratio. Finally, this ratio is also sensitive to the statistical uncertainty in IGM modeling. We also note that several effects might diminish Ly$\alpha$ radiation in high-density environments \citep{Neufeld_1990}. Recently, \citet{Dijkstra_DCBH_2016} argued that collisional de-excitation plays an important role for Ly$\alpha$ signatures of DCBHs. Our simulations do not resolve such high densities but these effects could be incorporated into the input spectrum, e.g. with a broader Ly$\alpha$ line. Although the \HeII/Ly$\alpha$ line ratio is affected, we do not expect significant changes to the Ly$\alpha$ spectral profile or surface brightness profile.

Perhaps the most important caveat is that although we accurately incorporate Ly$\alpha$ trapping, ionizing radiation, chemistry, and cooling into our hydrodynamical study we have done so in spherical symmetry. The one-dimensional approximation provides significant insight into the CR7 source. However, three-dimensional effects are likely to be important. For example, geometric effects such as gas clumping, rotation, filamentary structure, and anisotropic emission from the source often lead to anisotropic escape, photon leakage, or otherwise altered dynamical impact. This may increase the amount of residual \HI\ within the galaxy and perhaps more naturally reproduce the observed velocity offset. One might consider adding a layer of complexity with semi-analytic prescriptions within our one-dimensional models, e.g. self-shielding clouds or low column density holes. However such alterations would no longer be hydrodynamically self-consistent. Although three-dimensional hydrodynamical simulations exist they do not incorporate fully-coupled Ly$\alpha$ feedback. Post-processing studies of Ly$\alpha$ radiative transfer in similar contexts provides insight about the impact of three-dimensional effects \citep{Dijkstra_Kramer_2012,Behrens_2014,Duval_2014,Zheng_Wallace_2014,Smith_2015}. Ly$\alpha$ observables are certainly affected by viewing angle with higher equivalent widths, escape fractions, and line-of-sight fluxes for face-on inclinations than edge-on directions, e.g. due to the formation of a disk or jet. Three-dimensional simulations also demonstrate the importance of small-scale structure in the interstellar medium. Both of these effects could change the \HeII/Ly$\alpha$ line ratio by a factor of a few. Additional uncertainties in the galaxy model parameters and initial conditions may produce different observables. For example, Ly$\alpha$ feedback and radiative transfer observables may be affected by altering the gas density profile. Our choice of an isothermal profile is an idealized setup that may affect the early gas dynamics but matters less as the system evolves, especially in relation to the expanding shell. Our model uncertainties are likely dominated by our overall assumption of a one-dimensional geometry as opposed to the details of this one-dimensional profile. Still, in the context of the CR7 galaxy the assumption of spherical symmetry ought to provide a qualitatively correct, first-order, prediction of the Ly$\alpha$ signature.

%% file: figures/rv/caption.tex
Radial position~$r_\text{sh}$ and velocity~$v_\text{sh}$ of the expanding shell for each model. The Pop~III and MBH models are shown by the solid black and red curves, while the Boost and Leak modifications are denoted by dashed and dotted curves of the corresponding color. The grey dash-dotted curves are fits using Equation~(\ref{eq:shell_sol}), only provided for the Boost models to avoid overcrowding the figure. For reference, $\beta = 0.26~(0.15)$ for the Pop~III (MBH) Boost case. The thin curves represent simulations without Ly$\alpha$ coupling, demonstrating that Ly$\alpha$ radiation pressure is dynamically important. The radius was selected based on the peak number density within the shell.

%% file: figures/hydro/caption.tex
Evolution of the gas number density~$n$ and velocity~$v$ for $t = \{1, 2, 3, 5, 8.5, 20, 60\}$~Myr for the MBH model. The dotted curves are from a higher resolution simulation while the dashed curves are from a lower resolution simulation. The shell outflow structure is the predominant feature throughout the hydrodynamical simulation. The other models also exhibit qualitatively similar radial profiles.

%% file: figures/acceleration/caption.tex
Radial profile of the acceleration due to Ly$\alpha$ photons~$a_\alpha$, gas pressure~$a_\text{P}$, ionizing radiation~$a_\gamma$, and gravity~$a_\text{grav}$ (dark matter + baryons). Both the Pop~III (upper) and MBH (lower) models are shown 3~Myr into the simulation. Ly$\alpha$ trapping can be quite significant. We note that $a_\text{P} < 0$ inside the shell due to a negative pressure gradient (shown as dashed lines).

%% file: figures/flux/caption.tex
Observed line-of-sight flux as a function of Doppler velocity $\Delta v = c \Delta \lambda / \lambda$ for the Pop~III and MBH models. The grey curves are intrinsic profiles while the black and red curves represent a plausible reprocessing due to scattering in the IGM based on the model from \citet{Laursen_2011} and $\tau_\text{GP}^\text{red}$ from Equ.~(\ref{eq:tau_GP}), respectively. These are shown for early to late times of $0.5$~Myr (dotted), $3$~Myr (solid), and $8$~Myr (dashed). The observed velocity offset of +160~km\,s$^{-1}$ (vertical line) is roughly reproduced in the MBH case but not in the Pop~III model. The uncertainty due to DEIMOS/Keck and X-SHOOTER/VLT spectral resolution is illustrated by the grey vertical regions (see Section~\ref{subsec:basic_observational_properties} for additional details).

%% file: figures/v_offset/caption.tex
The time evolution of the velocity offset for the red peak of the intrinsic line-of-sight flux. The Pop~III and MBH models are shown by the solid black and red curves, while the Boost modification is denoted by dashed curves of the corresponding color. The location of the offset is fairly constant throughout the simulations.

%% file: figures/SB_r/caption.tex
Evolution of the radial surface brightness profile for the Pop~III and MBH models at $t = \{1, 2, 3, 5, 8.5, 60\}$~Myr. The late dashed curve is from a simulation with lower spatial resolution. For the intrinsic grey curves the region is more uniform and extended in the Pop~III case while the MBH emission more closely follows the shell expansion as a result of increased Ly$\alpha$ trapping. The black and red curves are corrected for transmission through the IGM based on the model from \citet{Laursen_2011} and $\tau_\text{GP}^\text{red}$ from Equ.~(\ref{eq:tau_GP}), respectively. After scattering in the IGM the Pop~III model appears much more compact whereas the MBH model remains spatially extended.

%% file: figures/SB_rv/caption.tex
The spectral line intensity as a function of radius for the MBH model at early ($t = 3$~Myr) and late ($t = 60$~Myr) times. In the ensuing time the shell radius has expanded from $r_\text{sh} \approx 0.5$~kpc to $r_\text{sh} \approx 4$~kpc, which makes a visible difference in the extent of the Ly$\alpha$ emitting region. The images are corrected for transmission through the IGM where the upper and lower panels are based on the model from \citet{Laursen_2011} and $\tau_\text{GP}^\text{red}$ from Equ.~(\ref{eq:tau_GP}), respectively.

%% file: text/Conclusions.tex
\section{Summary and Conclusions}
\label{sec:conclusion}
Ly$\alpha$ emitting sources provide intriguing hints about the formation and evolution of galaxies in the high-redshift universe. The CR7 source at $z \approx 6.6$ presents a unique opportunity in this regard due to its exceptionally bright Ly$\alpha$ and \HeII\ 1640~\AA\ line emission but absence of metal lines. Previous investigations have considered the possibility that we are witnessing a young primordial starburst or direct collapse black hole. In this work we examine the connection between the emission source and Ly$\alpha$ observables. Specifically, \citet{Sobral_2015} report a +160~km\,s$^{-1}$ velocity offset between the Ly$\alpha$ and \HeII\ line peaks which we self-consistently reproduce in one-dimensional radiation-hydrodynamics simulations. Our simulations represent the first hydrodynamical study incorporating accurate Monte-Carlo radiative transfer calculations of Ly$\alpha$ radiation pressure. The CR7 galaxy is an ideal application as it allows direct comparison with current observations.

In this pristine environment the dominant physical processes are gravity, hydrodynamical gas pressure, and radiative feedback from Ly$\alpha$ photons as well as the thermal and chemical coupling of ionizing radiation. The source ionizes its own local bubble and drives an expanding shell of gas from the centre. The details depend on the properties of the source spectrum which we categorize as either a Pop~III star cluster with $10^5$~K blackbody emission or a massive black hole with a nonthermal Compton-thick spectrum. Our results may be summarized as follows:
\begin{itemize}
  \item[(1)] The MBH model results in greater photon trapping within the shell because the harder spectrum reduces the average ionization cross-section for hydrogen yielding extra residual \HI.
  \item[(2)] The extent of the Ly$\alpha$ emitting region is correlated with the shell radius implying a likely source lifetime of $10-100$~Myr.
  \item[(3)] The MBH model reproduces the $+160~\text{km\,s}^{-1}$ velocity offset and the $16$~kpc region size while the Pop~III model does not.
\end{itemize}
We emphasize that these models represent a particular choice for the spectral energy distribution and should be viewed as such. Nonetheless, our results provide evidence corroborating a direct collapse black hole scenario. Indeed, we have presented independent and complementary evidence that CR7 hosts a DCBH, next to arguments based on metal enrichment which assesses the likelihood of forming Pop~III stars around $z \approx 6.6$ in massive host systems \citep[e.g.][]{Hartwig_2015}. It is known that a strong background of Lyman-Werner radiation is able to suppress star formation until the buildup of atomic cooling haloes, i.e. systems with virial temperature $T_\text{vir} \gtrsim 10^4$~K such that Ly$\alpha$ cooling is enabled \citep[e.g.][]{Haiman_LW_1997}. Furthermore, \citet{Visbal_CR7_2016} demonstrate that if local reionization occurs early enough and with sufficient ionizing flux the photoheated gas around a $M_\text{vir} \sim 10^9~\Msun$ halo with star-free progenitors has a cosmological Jeans mass that is comparably large. In such cases the metal enrichment is considerably delayed and potentially results in a rapid Pop~III starburst with Ly$\alpha$ luminosity comparable to that of CR7. However, the Ly$\alpha$ signatures are still more easily explained by the DCBH scenario.

Finally, we briefly discuss additional observational signatures that may distinguish a DCBH from a Pop~III starburst. The formation of a DCBH is likely to generate rapid rotation, producing a jet. However, CR7 was not detected in the Chandra COSMOS Survey so its X-ray luminosity in the energy range $0.2 - 10$~keV is below $10^{44}~\text{erg~s}^{-1}$ \citep{Elvis_2009}. This is not surprising because such a large flux would indicate the jet is pointing directly towards us. At other angles the low disk temperature, high-$z$ suppression factor, and absorption in the Milky Way imply an even lower observed flux \citep{Hartwig_2015}, pushing the limits of even next-generation X-ray observatories, such as the X-ray Surveyor \citep{X-Ray-Surveyor}. On the other hand, a detection of radio emission would also provide strong evidence for the presence of a massive black hole. Future improved simulations are likely to further elucidate the nature of the remarkable CR7 source, allowing more robust predictions for its multi-wavelength signature.

\section*{Acknowledgements}
We thank the referee for constructive comments and suggestions which have improved the quality of this work. AS thanks the Institute for Theory and Computation (ITC) at the Harvard-Smithsonian Center for Astrophysics for hosting him as a visitor through the National Science Foundation Graduate Research Internship Program. This material is based upon work supported by a National Science Foundation Graduate Research Fellowship. VB acknowledges support from NSF grant AST-1413501, and AL from NSF grant AST-1312034. The authors acknowledge the Texas Advanced Computing Center~(TACC) at the University of Texas at Austin for providing HPC resources.